\documentclass[a4paper]{article}

\usepackage[english]{babel}
\usepackage[utf8x]{inputenc}
\usepackage[T1]{fontenc}

\usepackage[a4paper,top=3cm,bottom=2cm,left=3cm,right=3cm,marginparwidth=1.75cm]{geometry}

\usepackage{amsmath}
\usepackage{graphicx}
\usepackage[colorinlistoftodos]{todonotes}
\usepackage[colorlinks=true, allcolors=blue]{hyperref}

\usepackage{multirow}
\usepackage{xcolor,colortbl}
\definecolor{lightgray}{gray}{0.7}

\title{Precious Time: Understanding Social Stratification in the Knowledge Society Through Time Allocation}
\author{Michal Kakol, Radoslaw Nielek, Adam Wierzbicki}

\begin{document}
\maketitle

\begin{abstract}
The efficient use of available resources is a key factor in achieving success on both personal and organizational levels. One of the crucial resources in knowledge economy is time. The ability to force others to adapt to our schedule even if it harms their efficiency can be seen as an outcome of social stratification. The principal objective of this paper is to use time allocation to model and study the global efficiency of social stratification, and to reveal whether hierarchy is an emergent property. A multi-agent model with an evolving social network is used to verify our hypotheses. The network’s evolution is driven by the intensity of inter-agent communications, and the communications as such depend on the preferences and time resources of the communicating agents. The entire system is to be perceived as a metaphor of a social network of people regularly filling out agenda for their meetings for a period of time. The overall efficiency of the network of those scheduling agents is measured by the average utilization of the agent’s preferences to speak on specific subjects. The simulation results shed light on the effects of different scheduling methods, resource availabilities, and network evolution mechanisms on communication system efficiency. The non-stratified systems show better long-term efficiency. Moreover, in the long term hierarchy disappears in overwhelming majority of cases. Some exceptions are observed for cases where privileges are granted on the basis of node degree weighted by relationship intensities but only in the short term.
\end{abstract}

\section{Introduction}
Although most of us probably tire of endless meetings, in reality, a substantial part of added value produced by knowledge workers is somehow related to meetings. Passing or gathering information, evaluating progress of team members, assigning tasks, selling ideas or asking people for help—such activities require not only our time but also synchronization of people participating in meetings. The question as to who should synchronize with whom is determined by social hierarchy. If someone wants to sell you a new insurance policy, you can pick the meeting place. On the other hand, if you want to convince your boss or supervisor to allocate some resources for your project, you must adjust to his or her daily plans.

Some signs of hierarchy are visible in nearly all communities. Existing exceptions, such as Anonymous, hippie communes, or many anarchists, have taken their lack of hierarchy as their hallmark. A variety of processes can influence the order of hierarchy in a particular community. Sometimes, especially in companies, a formal title held by a person is translated almost directly into a position in the hierarchy. In other cases, individual trails 
or particular skills 
are more important. Sometimes, hierarchy is related to race, skin color, place of birth, or wealth. 

Hierarchy in a society results from a stratification system. Its components, e.g., social processes and institutions, generate inequalities in control over resources among its members  \cite{grusky1994contours}. The privileged enjoy a disproportionate share of valued resources, thereby constituting a hierarchy based on social traits. Eventually, as a feedback loop, the hierarchy recreates the inequalities. In this article, we focus on the existence of hierarchies in the knowledge society, where time and exchange of ideas during (virtual or real, but equally time-consuming) meetings are a basis of work and added value creation. In its most developed form, the knowledge society is exemplified by the Silicon Valley. The motto of this paper is a quote from Vivek Wadhwa, a well-known American technology entrepreneur and academic. The quote inspires several of our hypotheses, which are related to a decreased effectiveness and stability of hierarchies in the modern knowledge society.


The relationship between time and social position in the knowledge society seems evident, but is still not sufficiently understood. Broadly speaking, agents’ behaviors and privileges influence their ability to manage time (described in detail in the Model section). Insight into the use of time, a key economic component affected by social stratification, can have practical implications. Productivity of a company can be determined largely by rigidness of organizational hierarchy and enforced strategies of managing workers’ time. This is especially true for companies profiting from the knowledge economy. As a matter of fact, decreasing hierarchies and reducing constraints on individual time management has been advocated as a successful recipe for knowledge companies \cite{tapscott2008wikinomics}.



Our approach assumes that agents do not care about other agents’ utilities or performance on the societal level. Agents are greedy, and attempt to maximize only their own utility. However, we are concerned with the global efficiency of a social system. Measuring the efficiency of social systems is a difficult task that can be addressed from various perspectives. In our research, we use a social simulation approach that enables us to investigate the emergent effect of individual agent's strategies on the global welfare.


The following hypotheses, regarding our model of the knowledge society, are tested in this article:
\begin{itemize}
\item \textbf{H1:} Egalitarian knowledge societies and communities are more efficient;
\item \textbf{H2:} Application of advanced procedures of time allocation (more intelligent algorithms) on the agent level helps to overcome the influence of stratification in knowledge society;
\item \textbf{H3:} Social stratification is an emergent property, but does not persist in the long term in stable knowledge societies.
\end{itemize}


In section Related works, we present related work in sociology and, specifically, in stratification, including related work on its modeling. Later sections cover the description of our multi-agent model (section \textit{Model}), the measures used to evaluate it (section \textit{Measures}), and the results in the order of tested hypotheses (section \textit{Results}). Finally, we conclude by discussing the results and future work in \textit{Discussion}.

\section*{Related works}

It is widely agreed that differentiation of social groups into hierarchical layers is present in all developed societies. It is a worldwide phenomenon and except for utopian visions, no fully egalitarian social system has yet been found to work. However, since the underlying reasons can be disputed or interpreted through the lens of ideology, we venture to relate a couple of examples. These inequalities are so deeply embedded in our culture that not only do the privileged legitimize the mistreatment of some groups, but paradoxically the mistreated also accept and justify this mistreatment. \cite{jost_decade_2004} Parents show us our position in society and teach us to expect that same level. A study by \cite{hurwitz_impact_2011} has shown that legacy students (whose parents attended a particular college) are more likely to be admitted to elite colleges. We expect this, and we obtain it, according to the saying that “the rich get richer.” \cite{mulder_intergenerational_2009} found that today’s technological advances make it easier to transfer wealth, and thus also the associated status, between generations.  

Are inequalities that bad after all? The evidence is mixed and often ideologically underpinned. \cite{wilkinson2009spirit}, with a self-explaining title, “Why more equal societies almost always do better,” linked inequality with various social problems (e.g., obesity, homicide, and higher infant mortality). Despite the existence of sound evidence confirming that link, there is an ongoing discussion regarding the study’s methodology. \cite{snowdon_spirit_2010} \cite{saunders_beware_2010} In contrast to the mentioned findings, \cite{rowlingson_does_2011} attempted to prove that no correlation exists between income equality and social problems. \cite{atkinson_public_1996} went even further and claimed that inequality fuels economic growth. Although the issue of social stratification and inequality has been a subject of extensive research and debate over many years, the relation between inequality and prosperity still remains unsettled.

Stratification is most often considered in terms of economic inequality. Nevertheless, let us consider a hypothetical example in which the president of the United States finds an exceptionally urgent matter to discuss with the author of this article. Even under such conditions, the invited author would obviously need to adapt to the president’s schedule. This example depicts well the obvious link between time availability and social status. 

Time usage and its availability can also be a consequence of stratification – those on the bottom of a hierarchy have less control over how they use their time and must adapt to those at the top of the ladder. One can argue that there is hardly any relation, but since economics in general is the study of how agents allocate scarce resources, time allocation is of economic interest, as it is one of the key factors in understanding economic phenomena. Moreover, time is a component of nearly every economic undertaking. It is a part of market labor and a key input to consumption, as consumed commodities are produced with it, as much as market goods and production technology and the consumer’s time \cite{becker1965theory}. The classic literature on time allocation in the United States can be found in \cite{ghez1975allocation} and \cite{juster_time_1985}, which examined trends of time allocation based on time diaries, and an extension of it can be found in \cite{robinson_time_2010}. Moreover, \cite{aguiar_measuring_2006,aguiar_increase_2008} reported an interesting link between time usage patterns and socio-economic status according to which, over the last 20 years, less educated men increased the time they allocated to leisure, while more educated men reported a decrease.  

To the authors’ best knowledge, there are several computational models of societies incorporating hierarchy that focus more or less directly on stratification. \cite{small_finding_1999} proposed a model of Tongan society that simulated the relationship of warfare and traditional Tongan rules of marriage. Another model of Pacific island societies, proposed by  \cite{younger_leadership_2010}, modeled the relationship between limited resource usage and different leadership models. \cite{robison-cox_simulating_2007} introduced a simulation of promotion competition between male and female workers in a corporate environment, specifically modeling the disparity of female representation at the highest corporate management level. Although the proposed simulation included the temporal dimension, it modeled the problem on a company level, and its resolution could not cover workers’ social relations or schedules.

\cite{fischer2007evolutionary} described a model of time-use behavior. The model was fed with an empirical dataset gathered in a time-use diaries survey, but ignored the social aspect of time-allocation and the role of interaction between agents. Models in which interactions between agents are used for time allocation have been researched by \cite{wawer_patterns_2009}, but for different purposes, namely emergence of time-usage patterns  and infection spreading \cite{nielek_two_2008}. \cite{sutcliffe_computational_2012} also introduced a simulation of the emergence of complex social networks based on the intensity of relations and provided an overview of similar works, i.e., stratification into layers on an individual level by intensity of relations. The approach presented in the aforementioned paper has been implemented in our model. 

\section*{Model}\label{model}

The problem we have modeled consists of two components, the relation between the agents and the communication occurring between them, making it not only a social network but also a communication system. Formally, we move the problem to a social space with a defined topology, with points representing agents and metrics of their distance, e.g., their relation strength (i.e. social network). The relations of the actors can potentially be represented as vectors and the network as a summation of these vectors, i.e., as a two-dimensional matrix. Such matrices in \cite{kluver_dynamics_2000} are called socio-matrices or Moreno matrices; moreover,  \cite{leydesdorff_evolution_1994} introduces an analogical formulation for communication systems. Communication is modeled by an algorithm that searches for a communication partner based on the agent relation topology.

\begin{figure}[!b]
  \begin{center}
    \includegraphics[width=4in]{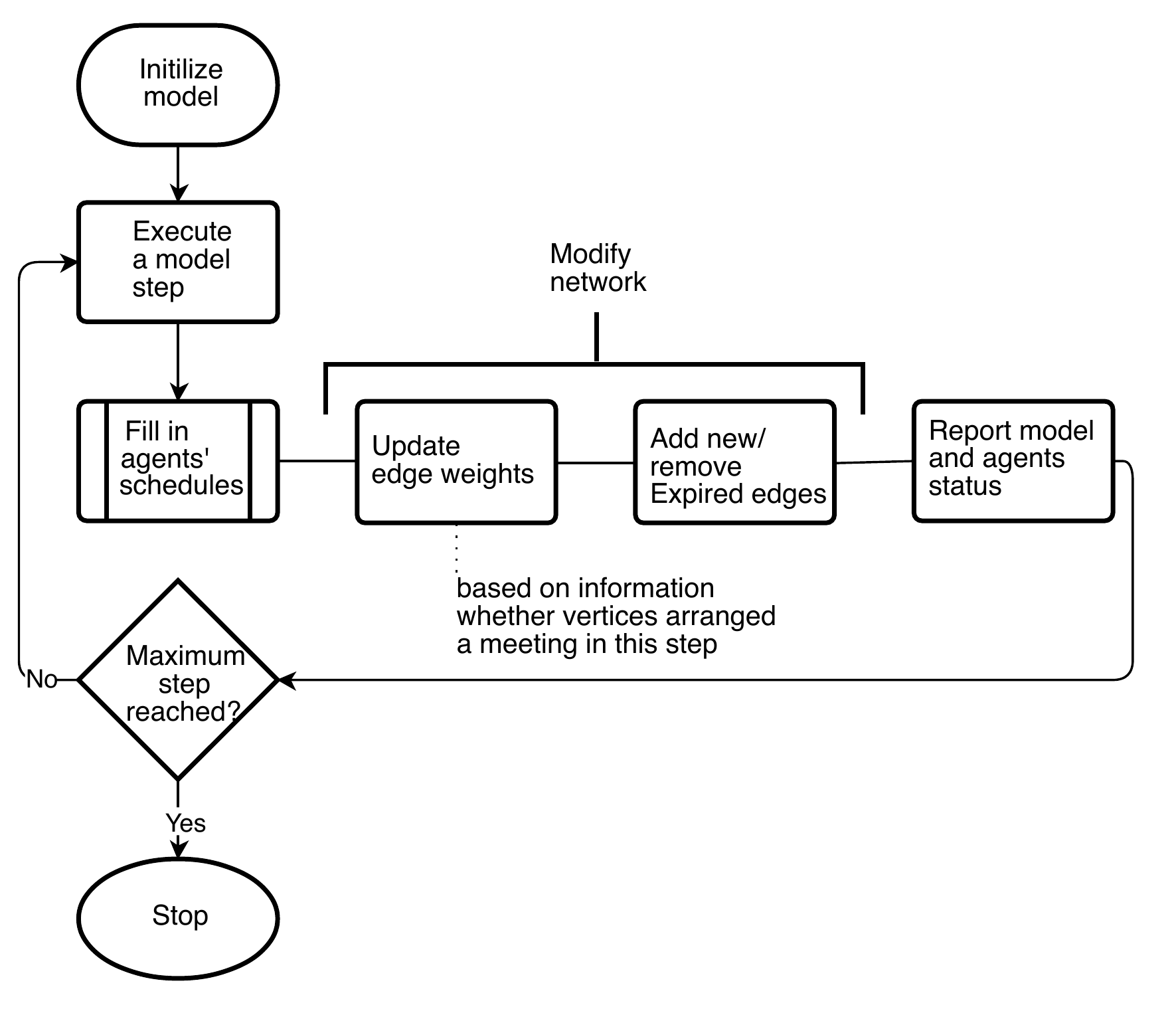}
  \end{center}
  
  \caption{\small Simulation loop elements.}
  \label{fig1}
\end{figure}

The model described in this paper is a multi-agent system implemented in NetLogo and based purely on a social network approach (no spatial dimension). The agents are embedded in a network of relations between them, communicating with each other to fulfill their need to “talk” on various subjects. The flowchart in Fig. \ref{fig1} depicts a single model step in which the agents are executed one after another (in a particular order) trying to fill in their “weekly” schedules. The details on each component of the model step are presented in the following sections. Note here that the order in which agents fill in their schedules may be fixed at the beginning of the simulation, during model initialization (for the scenarios that model a strict hierarchy) or may be determined in the beginning of each model step (for scenarios that model an egalitarian or mobile approach to scheduling agent meetings).

The size of the agent population is constant over the simulation. Removing or introducing new agents is not implemented in this model, and thus a single simulation can be referred to as presenting the population in a short time frame, sufficiently short to say that the population does not change. A single agent corresponds to a person, and the edge between agents represents a symmetrical acquaintanceship – both agents are linked with equal strength, and therefore both are equally capable of arranging a meeting. 

According to the social brain hypothesis and the findings of Dunbar, human relationships are characterized by different levels of intensity \cite{dunbar_determinants_1996,dunbar_social_1998}, which in the model is a link’s strength attribute. The strength of the relationship between agents depends on the frequency of interactions between them and has been modeled according to the study of \cite{hays_day--day_1989} -– agents communicate only with the agents they are directly connected to, the unused connections get weaker until they break up, and those frequently used are strengthened by analogy. The procedure of edge strength modification is described in detail in the Network evolution section. 

On the population level, there is a finite set of all possible topics/interests parameterized in the model as culture. Every agent has a set of interests, a subset (equal size for all the agents) drawn from the culture with a probability following a normal distribution. Therefore, some subjects are quite rare in the population (e.g., Persian poetry) and some very popular (e.g., football). Additionally, every agent possesses a willingness attribute for each subject, which is assigned randomly from zero to one with a uniform distribution for every interest, thus representing an agent’s inclination for a given subject. An agent's willingness expresses any individual agents creativity and ability to contribute new ideas on a topic. Finally, every agent is equipped with the same number of time slots to arrange its meetings in.

The source code of the presented model, as well as a comprehensive list of model parameters, together with additional resources, can be found on the model’s webpage \url{http://users.pjwstk.edu.pl/~s8811/papers/stratification/}.

\subsection*{Meeting scheduling procedures}

The modeled agents’ principal objective is a kind of rational choice, namely, to fulfill maximally their need to communicate. This need, i.e., “willingness,” is an attribute of each subject in a set of interests of a particular agent and is reduced each time an interaction between neighboring agents sharing the same interest occurs. The payoff of their actions is the sum of the used instances of willingness, which they try to maximize. At every turn for every agent, a scheduling procedure is invoked that is responsible for finding partners for meetings and filling agents’ time slots. A meeting occurs when two agents agree on a particular common free time slot and subject. Every scheduled meeting consumes a time slot and willingness to discuss a particular subject. Letting \( W_{s1}^A\) and \( W_{s1}^B\) be willingness to discuss a subject 1 for agents A and B, respectively, the remaining willingness for agent A after the meeting can be calculated using the following formula:

\begin{equation}\label{eq:eq1} 
{W'}_{s1}^A = {W'}_{s1}^A - min\left({W'}_{s1}^A, {W'}_{s1}^B\right)
\end{equation}

Notice that the result of eq. \ref{eq:eq1} for one of the two agents participating in the meeting will always equal zero, thereby excluding this agent as a communication partner for other agents in the current step. A meeting consumes a single time slot, with a length that is limited and might be interpreted as days of the week planned in advance by each agent.

The egoistic goal of the agent is to maximize its own willingness usage; hence, we use an average value of used willingness as an overall measure of network efficiency in the model, i.e., the extent to which the agents are utilizing their preferences. The average willingness usage expresses the global utilization of ideas generated by all agents in a knowledge society or community. Maximizing average willingness usage is a combinatory problem and can be solved in many ways, but finding an optimal solution requires algorithms with exponential complexity. As users have only limited resources, they must rely only on approximation algorithms (heuristics). Two heuristics have been implemented and tested: 

\begin{enumerate}
\item Simple -- agent A chooses a random agent B from its acquaintances and if they share a free time slot, selects a subject S, in order to maximize \(min\left(W_{s1}^A, W_{s1}^B\right) \); every agent repeats this procedure a limited number of times (model’s parameter),
\item Intelligent -- agent A iterates over the list of its’ subjects sorted in descending order by willingness to talk; for each of the iterated subjects all days of the week are used to find a partner B with maximal willingness to talk; this heuristic can be summarized as "trying until success".
\end{enumerate}

For a detailed flowchart, see Appendix I.


\subsection*{Stratifying the system}

The stratification in the proposed model manifests in three agent initialization orders followed by schedule filling, simulating different levels of inequality and rigidity as defined in \cite{grusky1994contours}.
\begin{enumerate}
\item \textbf{Egalitarian} -- The \textit{baseline} used in the model is a random order of executing the agents and is treated as a metaphor of an egalitarian system with equal opportunities to make appointments; the inequalities and rigidity in such a system are low.  
\item \textbf{Hierarchical} -- In the \textit{hierarchical} scheduling method, each agent in the system has a fixed place in the hierarchy, and thus the agents are executed in the same order throughout the entire model run, at every turn. This can be referred to as the highest level of inequality and rigidity in the system. 
\item \textbf{Mobile} -- In the \textit{mobile} method, differentiation of the agents occurs at every turn. At each turn, the agents (nodes) are ordered according to their network degrees weighted by their relations’ strengths, resulting in a medium level of inequality and rigidity in the system. A node's degree is interpreted (in accordance with much research in social network theory) as the nodes social capital or influence, which determines the node's ability to schedule meetings ahead of less influencial nodes.
\end{enumerate}

\subsection*{Network evolution}
In the initial network, relations between the agents are generated with a given density (we arbitrarily used settings of 0.2\% or 0.6\%). However, the network changes based on the outcome of interactions between agents. The evolution of the network is shaped by three procedures that are invoked after every turn: 
\begin{enumerate}
\item \textbf{Edge strength updating} -- basically, this procedure looks up all edges in the model and updates their strengths according to the activity at the last turn, namely, the strength of every edge that has been used for a successful meeting arrangement is increased and the strengths of unused edges are decreased; two separate updating functions have been implemented, linear and logarithmic, binding the pace of relation change with their strength,
\item \textbf{Edge creation} -- according to the simulation parameters, new edges are created using two heuristics, triadic closure (\textit{“my friend’s friend”}) \cite{oyster_groups:_2000} approach, in which there is some probability that a new edge appears between two agents that have at least one common friend, and the \textit{“random meeting”} approach, which is simulated by the creation of an edge between two random agents in the population. However, an additional parameter in the model was added controlling the choice of a new neighbor. By default, the neighbors are chosen randomly with equal probabilities, but a preferential mode can be activated to choose more likely new neighbors of a greater network degree, i.e., preferring more popular agents.
\item \textbf{Edge removing} -- edges that are not used sufficiently frequently (i.e., their strength reaches zero) are removed.
\end{enumerate}

\section*{Measures}

All measurements were taken after each turn, as all the agents’ schedules were filled and the network had changed accordingly. Along with the well-known network metrics (e.g., network density and average degree), other measures were also used - measures of ranking comparison and two measures expressing efficiency of different stratification procedures and heuristics for arranging meetings (all normalized to one).

\textbf{Schedule usage} represents the percentage of all days in the schedules used in one turn and is calculated using eq. \ref{eq2},

\begin{equation}\label{eq2} 
Schedule_{Usage}=\frac{\sum_{i=1}^NT_i}{N*weekdays}
\end{equation}

where \(N\) is the number of agents, and \(T_i\) is the number of used time slots of the \textit{i}-th agent. 

\textbf{Willingness usage} represents the percentage of “consumed” willingness of all agents in one turn in relation to all available willingness and is calculated using eq. \ref{eq3},

\begin{equation}\label{eq3} 
Willingness_{Usage}=\frac{\sum_{i=1}^NSU_i}{\sum_{i=1}^NS_i}
\end{equation}

where \(N\) is the number of agents, \(SU_i\) is the sum of used willingness points in the \textit{i}-th agent’s schedule, and \(S_i\) is the sum of willingness points of all subjects in the ith agent’s interest set.

\textbf{Kendall Tau distance}

To measure the differences between agents’ execution order lists for consecutive simulation steps, we arbitrarily decided to use the Kendall Tau distance. This distance is a measure between two full rankings (over the same set of n elements) that counts the number of pairwise disagreements between those two rankings \cite{Bansal_computing_2009}. We used the normalized Kendall Tau distance, i.e., multiplied by the maximum value of the metric: $\frac{n(n-2)}{n}$.

\begin{equation}\label{eq4} 
K\left(\sigma_1,\sigma_2\right)=\sum_{\left \{i,j\right \}\in P}\overline{K}_{i,j}\left(\sigma_1, \sigma_2\right)
\end{equation}

\begin{equation}\label{eq5} 
K_N\left(\sigma_1,\sigma_2\right)=\frac{2}{n\left(n-2\right)}K\left(\sigma_1, \sigma_2\right)
\end{equation}

where \(\sigma\) is a list or ranking, \(n\) is the number of elements in the list, \(P\) is a set of unordered pairs of distinct list elements, and \(\overline{K}_{i,j}\left(\sigma_1, \sigma_2\right)\) is a comparison of single elements \(i\) and \(j\) in both compared lists:

\begin{equation}\label{eq6} 
\overline{K}_{i,j}\left(\sigma_1, \sigma_2\right)=\begin{cases}0 \Leftrightarrow i,j & \textrm{are in the same order}\\1 \Leftrightarrow i,j & \textrm{are in reverse order}\end{cases}
\end{equation}

\textbf{Order list cardinality}
We introduced this measure to monitor the stability of the set of privileged agents in consecutive model steps, i.e., the agents executed first. Precisely, we used the top 100 agents from the order list representing the execution order of all 1,000 agents. To measure the differences between those incomplete rankings, we used the simplest measure available. In favor of distance measures for partial rankings \cite{Bansal_computing_2009}, we used the cardinality of the intersection of the partial rankings, thus monitoring the number of common agents remaining in the “elite” in consecutive steps.

\begin{equation}\label{eq7} 
\rvert \sigma_1\cap \sigma_2 \rvert
\end{equation}

\section*{Results}

\subsection*{Egalitarian societies are more efficient}
\begin{figure}[!b]
  \begin{center}
    \includegraphics[width=4.3in]{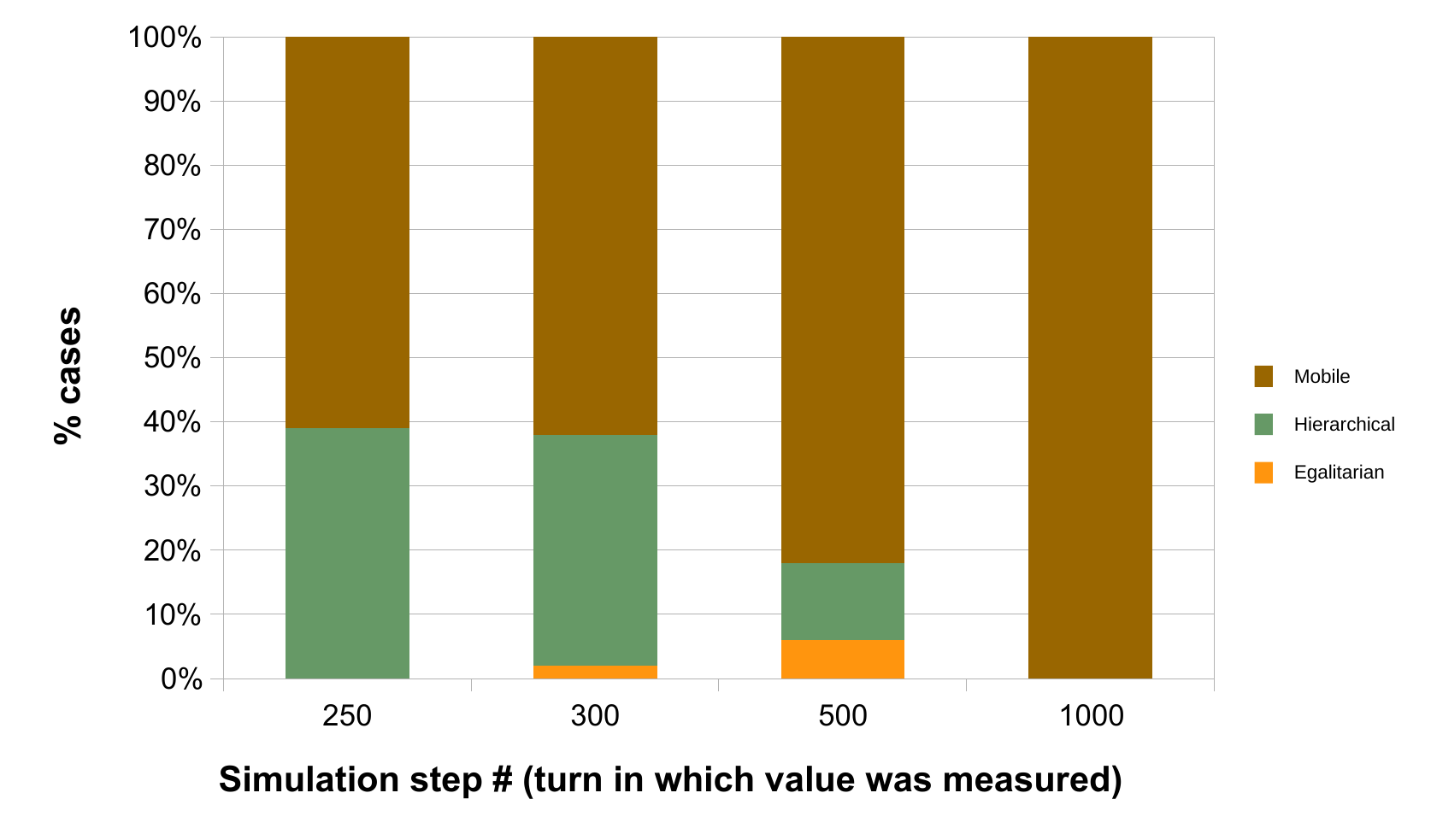}
  \end{center}
  
  \caption{\small Percentage of cases in which a parameter set results in the greatest Willingness Usage value for a particular Execution Order by simulation step number.}
  \label{fig2}
\end{figure}

The results of the 72 parameter combinations (four points in time for each parameter set are considered, i.e., the 200th step, 300th step, 500th step, and the last 1,000th step) prove that the Egalitarian execution method, which represents a system with no hierarchy at all, results in higher willingness usage in the majority of cases. The egalitarian scheduling method achieved the highest willingness usage mean in 76\% of tested runs; see Fig. \ref{fig2}. Most importantly, the pairwise comparison of all three stratification methods in all of the runs showed statistically significant differences in the means ($p<0.0001$, Bonferroni-Dunn test for multiple comparisons). In 19.1\% of the cases, a single compared pair of the mean values showed no significant difference. However, these were mostly pairs of Hierarchical and mobile (76.3\%). Moreover 74.5\% of the cases with an insignificantly different pair were those for which Egalitarian still showed the best performance in terms of average willingness usage value.


\begin{figure}[!b]
  \begin{center}
    \includegraphics[width=4.3in]{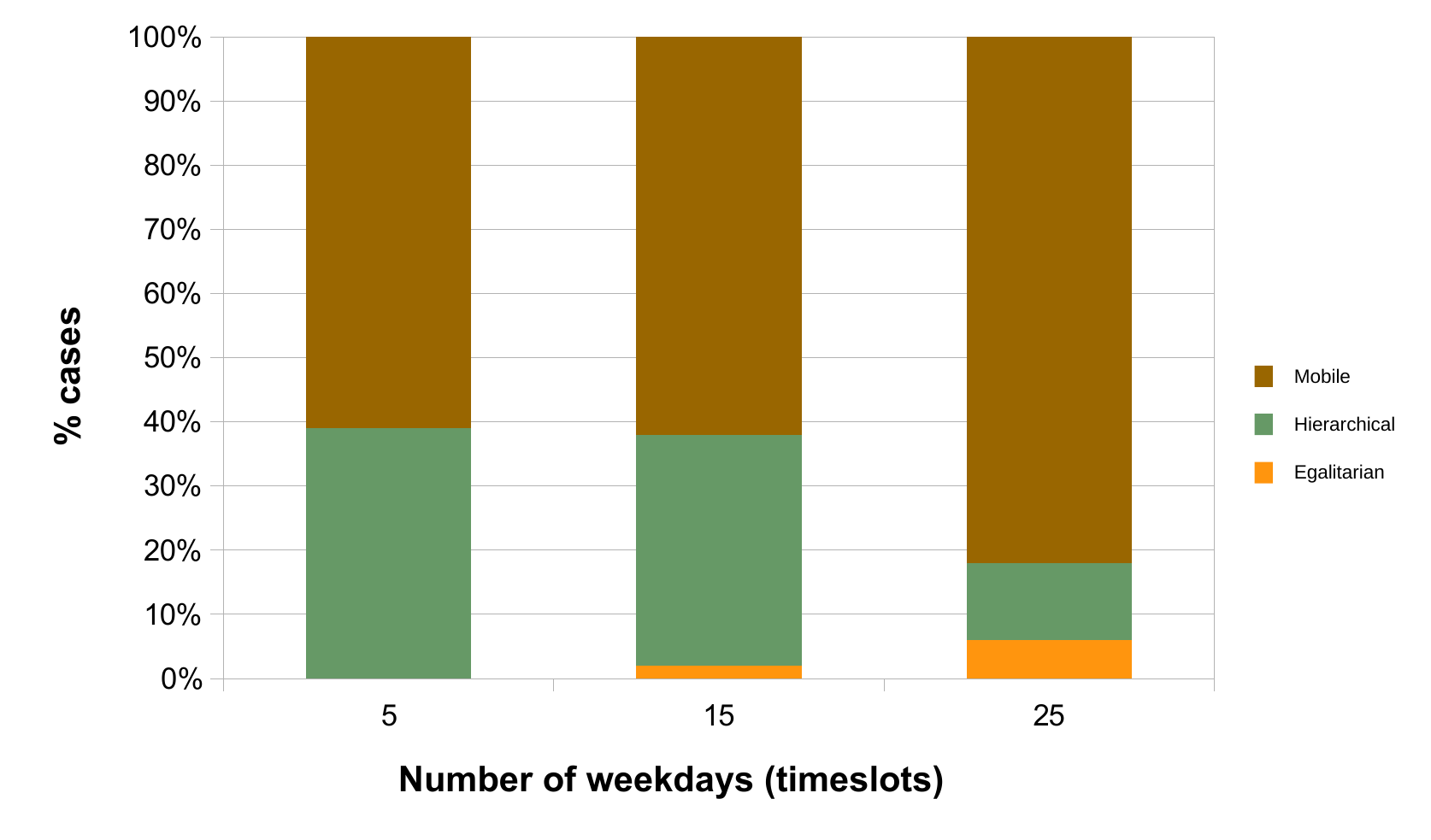}
  \end{center}
  
  \caption{\small Percentage of cases in which a parameter set results in the greatest Willingness Usage value for a particular Execution Order by number of time slots.}
  \label{fig3}
\end{figure}

Obviously, there are cases in which Egalitarian does not show a better performance than stratified mobile and Hierarchical. Apparently, the performance of the egalitarian system improves over time, as the random execution order shows the greatest willingness usage values in only approximately 60\% of cases for the first 300 turns, but 82\% for 500 turns, and eventually 100\% in the 1,000th turn (details on Fig. \ref{fig3}). This can be interpreted as indicating that random executions lead to the best long-term performance. Additionally, the advantage of random order in terms of willingness usage is most common for systems of scarce time resources (five time slots). The percentage of cases in which Egalitarian shows better efficiency is inversely proportional to the number of available time slots. Eventually, for 25 timeslots, these cases amount to only 68\%. 

The advantage of the privileged agents (first to be executed) is clearly visible in the data, and apparently this effect is common. The agents executed in first place surpass the rest of the agents in terms of willingness usage. The increase of efficiency on average for those first 10\% agents compared to the remainder 90\% of the system amounts to 20.3\%, 24.4\%, and 32.0\% for Egalitarian, Hierarchical, and mobile, respectively (calculated for all tested runs). This effect exists for all tested parameter sets, and under these conditions, the agents privileged by execution with higher priority are in the best position.

  

  

\begin{figure}[!b]
  \begin{center}
    \includegraphics[width=4.3in]{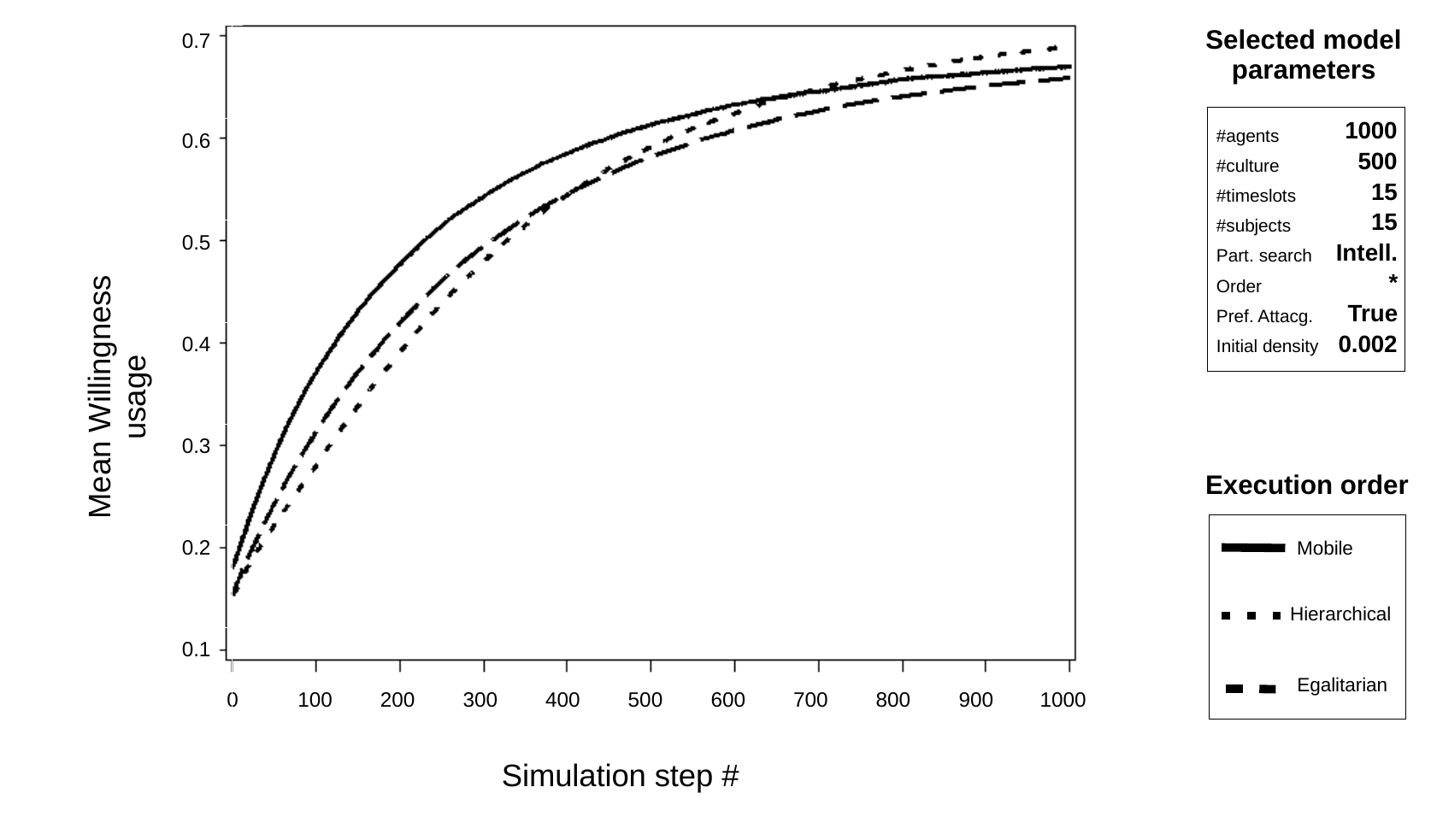}
  \end{center}
  
  \caption{\small Willingness usage values for the entire model run with the “Intelligent” partner finding heuristic.}
  \label{fig6}
\end{figure}

\begin{figure}[!b]
  \begin{center}
    \includegraphics[width=4.3in]{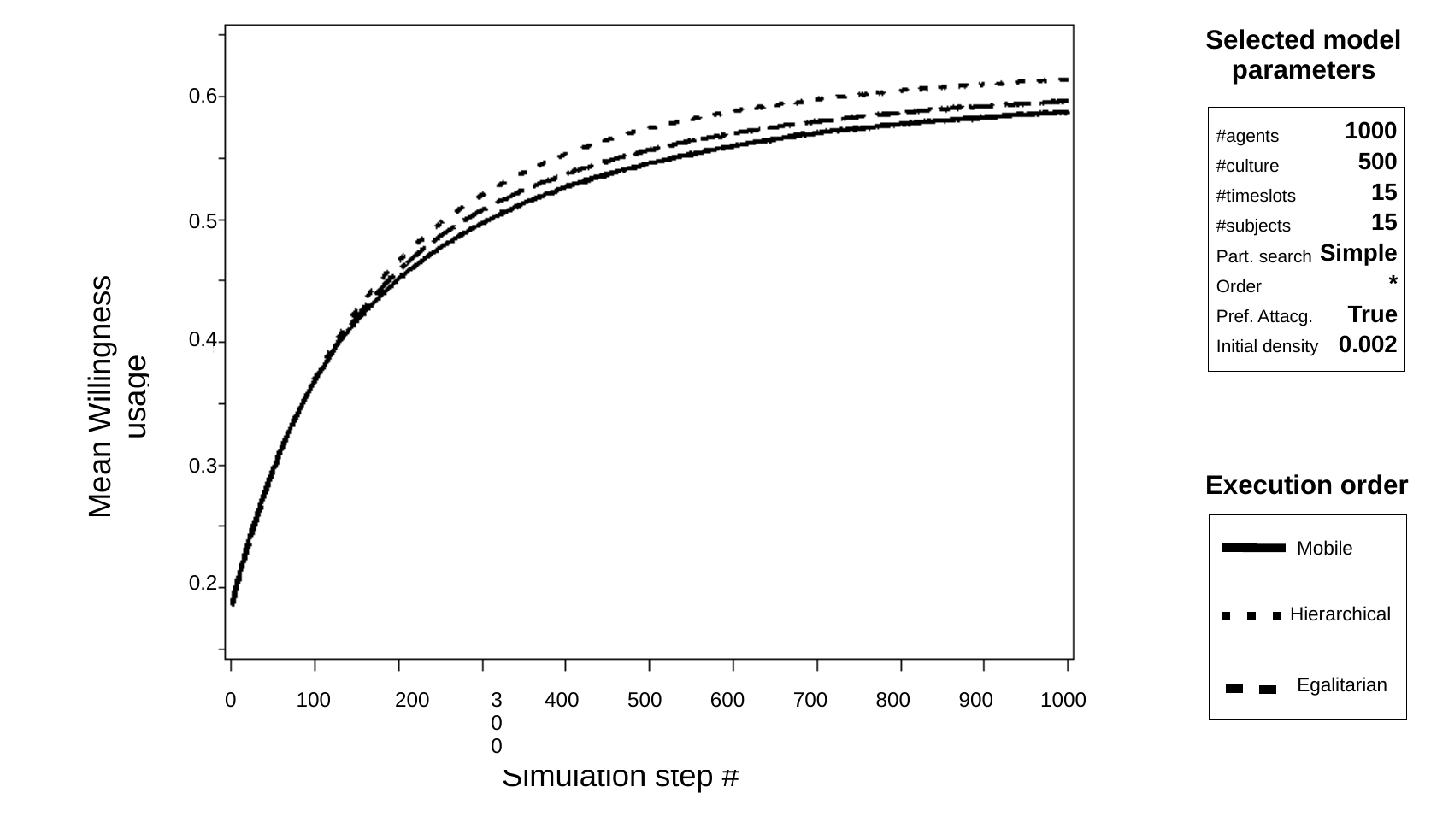}
  \end{center}
  
  \caption{\small Willingness usage values for the entire model run with the “Simple” partner finding heuristic.}
  \label{fig7}
\end{figure}

Fig. \ref{fig6} and \ref{fig7} depict the model willingness usage output for two time-allocation strategies, i.e., Intelligent and Simple. For both methods, the random execution Egalitarian shows better model efficiency in the area of model stability. For Intelligent time allocation, however, the stratified execution method by weighted degree has an advantage over random execution in some of its first hundred steps. This effect is temporary and involves only the warm-up phase. What happens exactly is that for the Intelligent partner finding heuristic, the Mobile method for ordering agents surpasses the Egalitarian one in terms of the willingness usage value. For a simple conversation arrangement scheme, Egalitarian resulted in the greatest willingness values in the overwhelming majority of cases. In contrast to what happens when Intelligent arrangements are used, the share of cases in which Egalitarian is the most efficient drops to roughly 22\% for the 250th step. However in the long term, that is at the 1,000th step, Egalitarian random execution again results in the best efficiency in 100\% of cases. 

The introduction of the Intelligent partner finding heuristic might be a metaphor for the digital revolution. The Intelligent heuristic leading to more sophisticated schemes of arranging meetings with other agents results in more efficient search for conversation partners. This can be referred easily to contemporary advances in communication, such as Internet usage, with which we are eventually going to find a “soul mate” sharing our interests. In light of the simulation results (see Table \ref{table1a_Simple} and Table \ref{table1b_Intelligent}), introduction of “intelligent” meeting arrangement schemes helps to boost a stratified system’s efficiency. However this effect applies to short-term results only as, in the long term, i.e., at approximately the 1,000th step, random execution approximating an egalitarian system shows the best efficiency after all. In other words, \textit{application of advanced procedures of time allocation at the agent level does not overcome the consequences of stratification.}

\begin{table}[!ht]
\caption{
{\bf Percentage of simulation runs in which particular execution method results in the highest willingness usage values for particular simulation step. Results for simple partner finding heuristic.}}
\begin{tabular}{l|l|r|r|r|r|}
\cline{2-6}
 & \textbf{Step} & \multicolumn{1}{c|}{\textbf{250}} & \multicolumn{1}{c}{\textbf{300}} & \multicolumn{1}{|c|}{\textbf{500}} & \multicolumn{1}{c|}{\textbf{1000}} \\ \hline
 \multicolumn{1}{ |c| }{\multirow{3}{*}{\textbf{Agentset switch}}} & Egalitarian &	100.00\% & 97.22\% & 94.44\% & 100.00\% \\ \cline{2-6}
\multicolumn{1}{ |c|  }{}  &	Mobile & 0.00\% &	0.00\% &	0.00\% &	0.00\%\\ \cline{2-6}
\multicolumn{1}{ |c|  }{} & Hierarchical	& 0.00\% &	2.78\% &	5.56\%	& 0.00\%\\ \hline
 
\end{tabular}
\label{table1a_Simple}
\end{table}

\begin{table}[!ht]
\caption{
{\bf Percentage of simulation runs in which particular execution method results in the highest willingness usage values for particular simulation step. Results for intelligent partner finding heuristic.}}
\begin{tabular}{l|l|r|r|r|r|}
\cline{2-6}
 & \textbf{Step} & \multicolumn{1}{c|}{\textbf{250}} & \multicolumn{1}{c}{\textbf{300}} & \multicolumn{1}{|c|}{\textbf{500}} & \multicolumn{1}{c|}{\textbf{1000}} \\ \hline
 \multicolumn{1}{ |c| }{\multirow{3}{*}{\textbf{Agentset switch}}} & Egalitarian &	22.22\% & 27.78\% & 69.44\% & 100.00\% \\ \cline{2-6}
\multicolumn{1}{ |c|  }{}  &	Mobile & 77.78\% &	72.22\% &	25.00\% &	0.00\%\\ \cline{2-6}
\multicolumn{1}{ |c|  }{} & Hierarchical	& 0.00\% &	0.00\% &	5.56\%	& 0.00\%\\ \hline
 
\end{tabular}
\label{table1b_Intelligent}
\end{table}


\subsection*{Emergence of social stratification}

While analyzing the model results, we wanted to know whether such loosening of system rigidity would lead towards randomness or back to strict order. To determine the answer, each model performed a dump of the agents’ identification number list in the order in which they were executed in the particular simulation step. We assume that comparing those execution order lists will provide the answer.

As described in the Measures section, the execution order list comparison involves a)computing the Kendall Tau distance between consecutive lists, and b)computing the order list cardinality of two consecutive execution order lists limited to only the first 100 agents. Note that the following section focuses on describing the results corresponding to the Mobile execution method only. Another reason for such a choice is the fact that distributions for the remainder of the execution methods are constant and can be summed up as follows. 

\begin{figure}[!b]
  \begin{center}
    \includegraphics[width=4.3in]{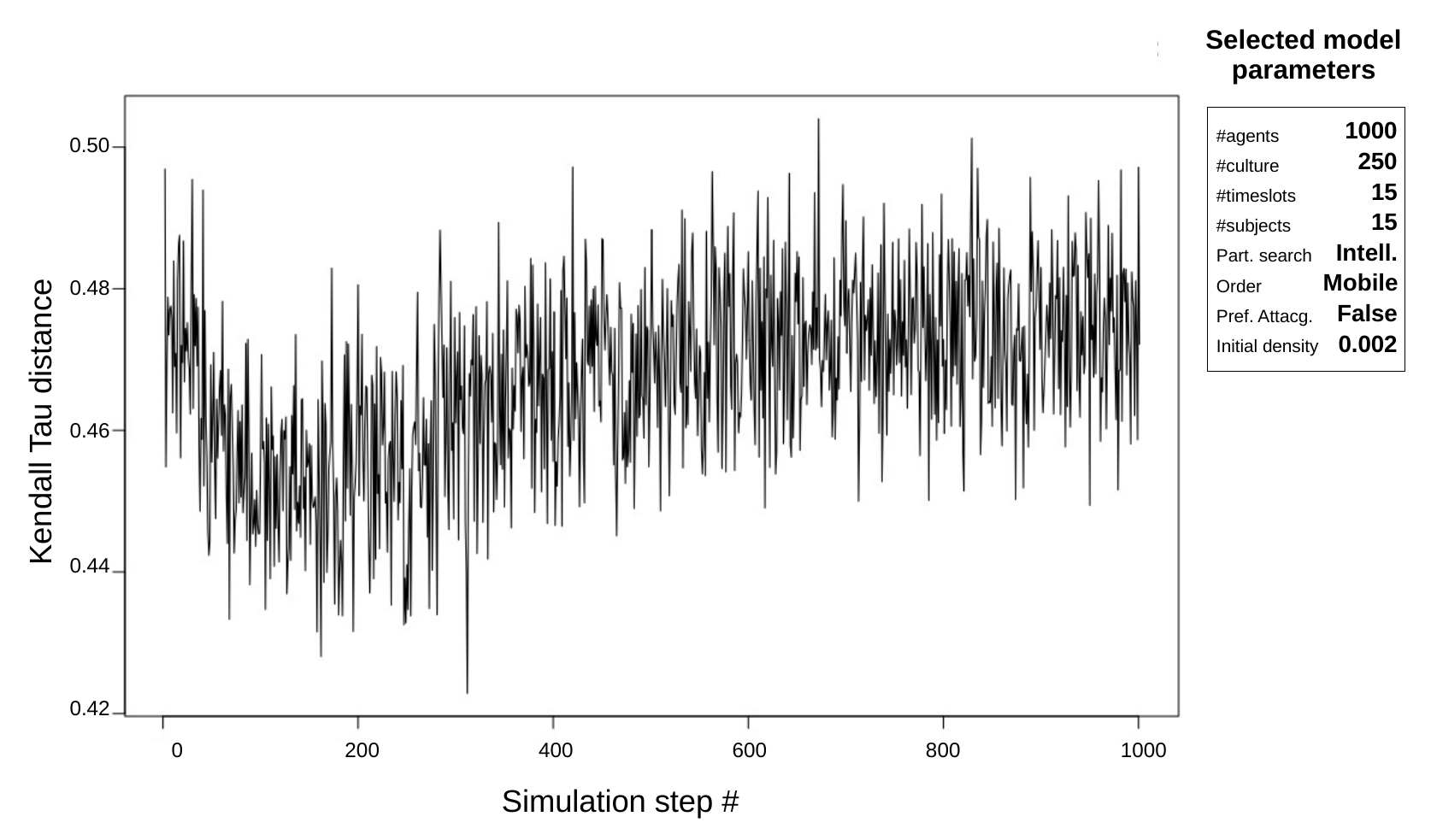}
  \end{center}
  
  \caption{\small Distribution of Kendall Tau distance for consecutive model steps for a sample set of parameters and Mobile execution only, raw data.}
  \label{fig8}
\end{figure}

\begin{figure}[!b]
  \begin{center}
    \includegraphics[width=4.3in]{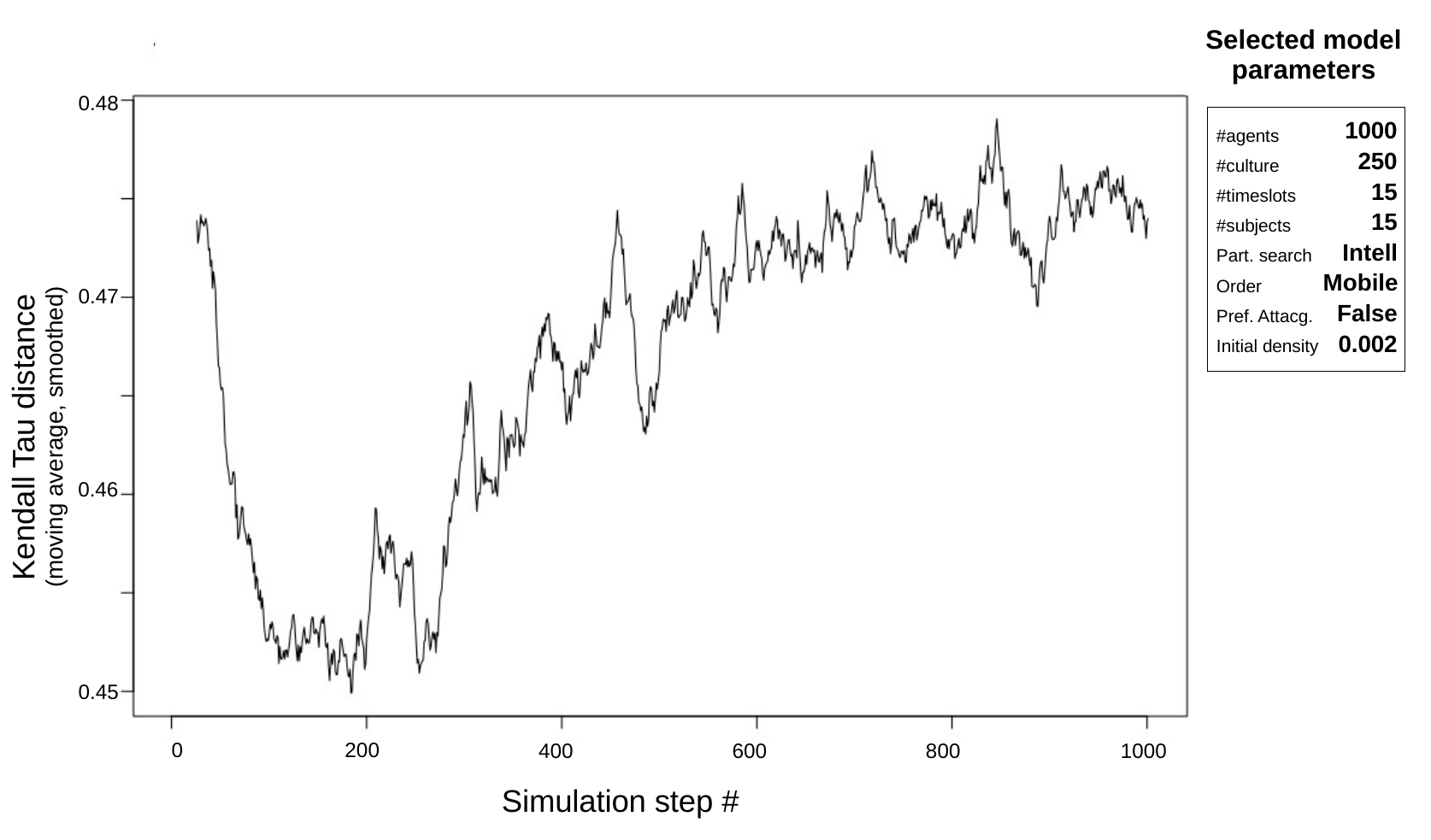}
  \end{center}
  
  \caption{\small Distribution of Kendall Tau distance for consecutive model steps for a sample set of parameters and Mobile execution only, moving average smoothing.}
  \label{fig9}
\end{figure}

In contrast to the results of the Egalitarian (random ordering) and hierarchical (fixed ordering) methods, the Kendall Tau distance distribution of Mobile shows more interesting features. Fig. \ref{fig8} and \ref{fig9} depict a typical run of the model and the values of the distance measure. The characteristic of most importance is the sharp decline in the first 200 to 300 steps and the steady climb afterwards. 

The plummeting of the Kendall Tau distance values in the warm-up phase of the model run exists in almost 67\% of tested cases. We take it as a sign of the progressive organization of the manner in which the agents are executed. Literally, the distance between compared consecutive order lists decreases over the warm-up phase. In other words, each consecutive list of agents’ execution order is more similar to the previous one, i.e., is more similarly ordered. After this decrease, the value of the distance climbs steadily, i.e., the consecutive order lists become less commonly ordered, and the system loosens its rigidity. This described upward trend exists in nearly 71\% of cases. 

\begin{figure}[!b]
  \begin{center}
    \includegraphics[width=4.3in]{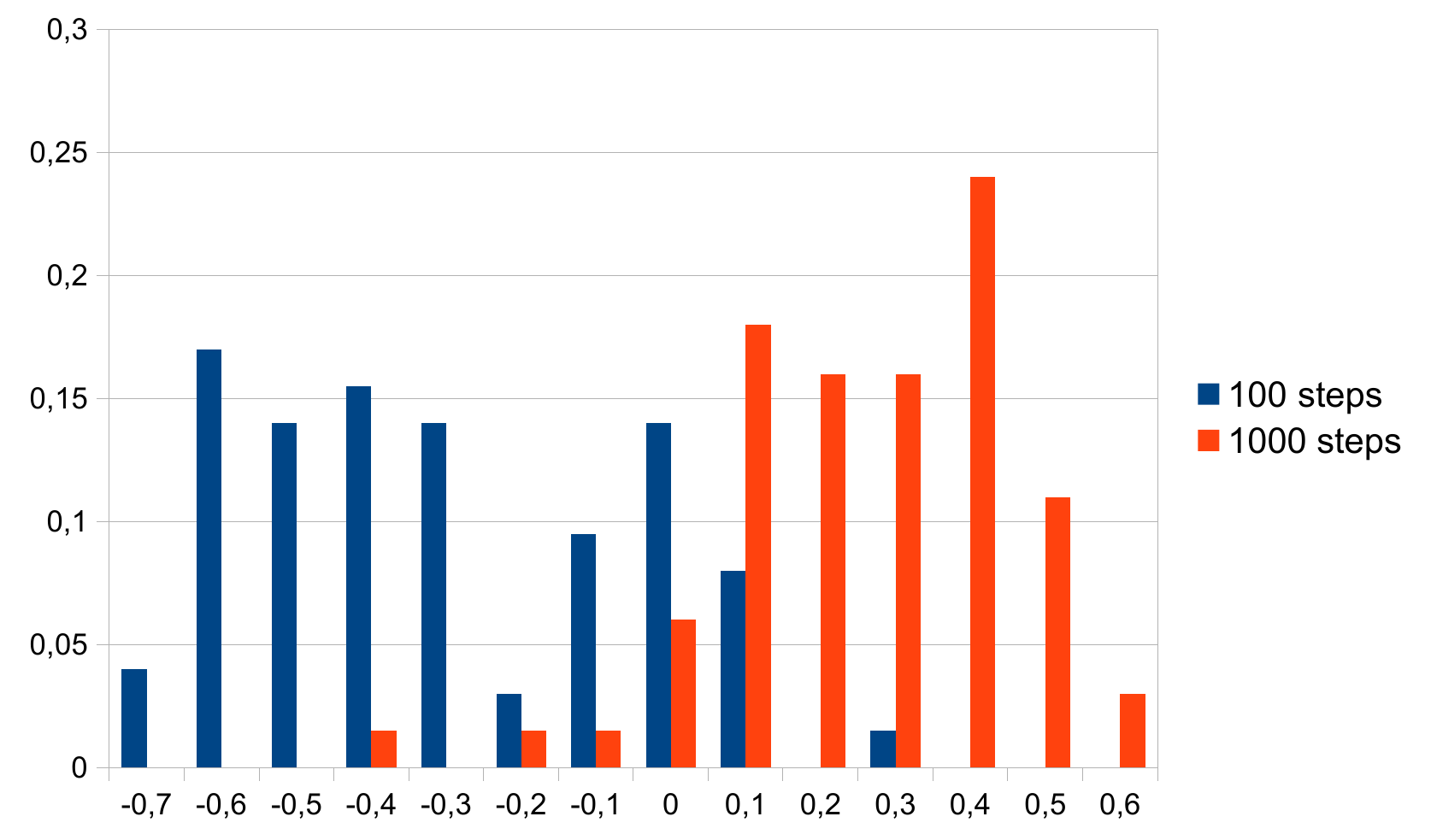}
  \end{center}
  
  \caption{\small Distribution of Kendall Tau distance-step number correlation values for the first 100 steps and an entire model run, for all tested parameter sets in general. Correlation values are on the x-axis.}
  \label{fig10}
\end{figure}

Apart from the visual identification of the trends, we measured the linear correlation between the distance value and the model step number. Tables \ref{table2a} and \ref{table2b} present the binned correlation values for different parameters separately, as well as for all the model runs. The advantage of negative correlation in the first steps and positive correlation in general over the entire model run is clearly visible. Also depicted in Figure \ref{fig10} is where the correlation value distribution is compared for the initial warm-up model phase and all 1,000 steps.

\begin{table}
\caption{
{\bf Summary of the Kendall Tau distance -- step number correlation for different parameter sets. Percentage of runs for given set of parameters for which correlation is within one of three ranges. Results for first 100 steps.}}
\begin{tabular}{cc|c|c|c|}
\cline{3-5}
 & & \multicolumn{3}{|c|}{\bf 100 steps} \\\cline{1-5}
\multicolumn{2}{ |c| }{ranges of correlation value} & $\langle-1;-0.1)$ &	$\langle-0.1;0.1\rangle$ &	$(0.1;1\rangle$\\ \hline \hline
\multicolumn{1}{ |c| }{\multirow{2}{*}{Culture size}}	& 250	& 66.67\% &	33.33\% &	0.00\%\\ \cline{2-5}
\multicolumn{1}{ |c| }{} & 500 & 66.67\% &	30.56\% &	2.78\%\\ \hline \hline
\multicolumn{1}{ |c| }{Number of} & 5 &	12.50\% &	83.33\% &	4.17\%\\ \cline{2-5}
\multicolumn{1}{ |c| }{weekdays} & 15 &	87.50\% &	12.50\% &	0.00\%\\ \cline{2-5}
\multicolumn{1}{ |c| }{(timeslots)} & 25 &	100.00\% &	0.00\% &	0.00\%\\ \hline \hline
\multicolumn{1}{ |c| }{New links} & 	Preferential &	66.67\% &	33.33\% &	0.00\%\\ \cline{2-5}
\multicolumn{1}{ |c| }{algorithm} & 	Uniform &	66.67\% &	30.56\% &	2.78\%\\ \hline \hline
\multicolumn{1}{ |c| }{\multirow{2}{*}{Arrangements}} & Intelligent	& 69.44\% &	30.56\% &	0.00\%\\ \cline{2-5}
\multicolumn{1}{ |c| }{} & Simple & 63.89\% &	33.33\% &	2.78\% \\ \cline{1-5} \cline{3-5}
& & \cellcolor{lightgray}\textbf{66.67}\% &	\cellcolor{lightgray}\textbf{31.94}\% &\cellcolor{lightgray}	\textbf{1.39}\%\\ \cline{3-5}
\end{tabular}
\label{table2a}
\end{table}

\begin{table}
\caption{
{\bf Summary of the Kendall Tau distance -- step number correlation for different parameter sets. Percentage of runs for given set of parameters for which correlation is within one of three ranges. Results for first 100 steps.}}
\begin{tabular}{cc|c|c|c|}
\cline{3-5}
 & & \multicolumn{3}{|c|}{\bf 1000 steps} \\\cline{1-5}
\multicolumn{2}{ |c| }{ranges of correlation value} & $\langle-1;-0.1)$ &	$\langle-0.1;0.1\rangle$ &	$(0.1;1\rangle$\\ \hline \hline
\multicolumn{1}{ |c| }{\multirow{2}{*}{Culture size}}	& 250	& 2.78\% &	25.00\% &	72.22\%\\ \cline{2-5}
\multicolumn{1}{ |c| }{} & 500 & 2.78\% &	27.78\% &	69.44\%\\ \hline \hline
\multicolumn{1}{ |c| }{Number of} & 5 &	0.00\% &	58.33\% &	41.67\%\\ \cline{2-5}
\multicolumn{1}{ |c| }{weekdays} & 15 &	0.00\% &	12.50\% &	87.50\%\\ \cline{2-5}
\multicolumn{1}{ |c| }{(timeslots)} & 25 &	8.33\% &	8.33\% &	83.33\%\\ \hline \hline
\multicolumn{1}{ |c| }{New links} & 	Preferential &	0.00\% &	22.22\% &	77.78\%\\ \cline{2-5}
\multicolumn{1}{ |c| }{algorithm} & 	Uniform &	5.56\% &	30.56\% &	63.89\%\\ \hline \hline
\multicolumn{1}{ |c| }{\multirow{2}{*}{Arrangements}} & Intelligent	& 0.00\% &	22.22\% &	77.78\%\\ \cline{2-5}
\multicolumn{1}{ |c| }{} & Simple & 5.56\% &	30.56\% &	63.89\% \\ \cline{1-5} \cline{3-5}
& & \cellcolor{lightgray}\textbf{2.78}\% &	\cellcolor{lightgray}\textbf{26.39}\% &\cellcolor{lightgray}	\textbf{70.83}\%\\ \cline{3-5}
\end{tabular}
\label{table2b}
\end{table}

An interesting pattern in changes of distance values can be observed for different numbers of available time slots. The discussed decrease in the initial model phase is more prominent when more time slots are available. For 25 weekdays, 100\% of cases are affected by the initial decrease, compared to only 12.5\% for five weekdays. A similar relation between number of weekdays and the cardinality of the intersection of two consecutive agent execution order lists can also be observed. 

The intersections of the consecutive order lists (first 100 agents only) are measured in order to examine whether the set of 10\% of agents to be executed first, the “privileged” caste, remains constant over the entire model run. The intersection cardinality value indicates the number of common agents in two consecutive order lists, e.g., 100: all agents the same in both lists, 70: 70 common agents in two lists. The common apparent effect is that the intersection value steadily decreases over the modeled 1,000 turns. Similarly to Kendall Tau distance, we measured the correlation between the intersection value and the model step number. The majority of tested parameter sets, viz., 72\%, reveal negative correlations.

\begin{figure}[!b]
  \begin{center}
    \includegraphics[width=4.3in]{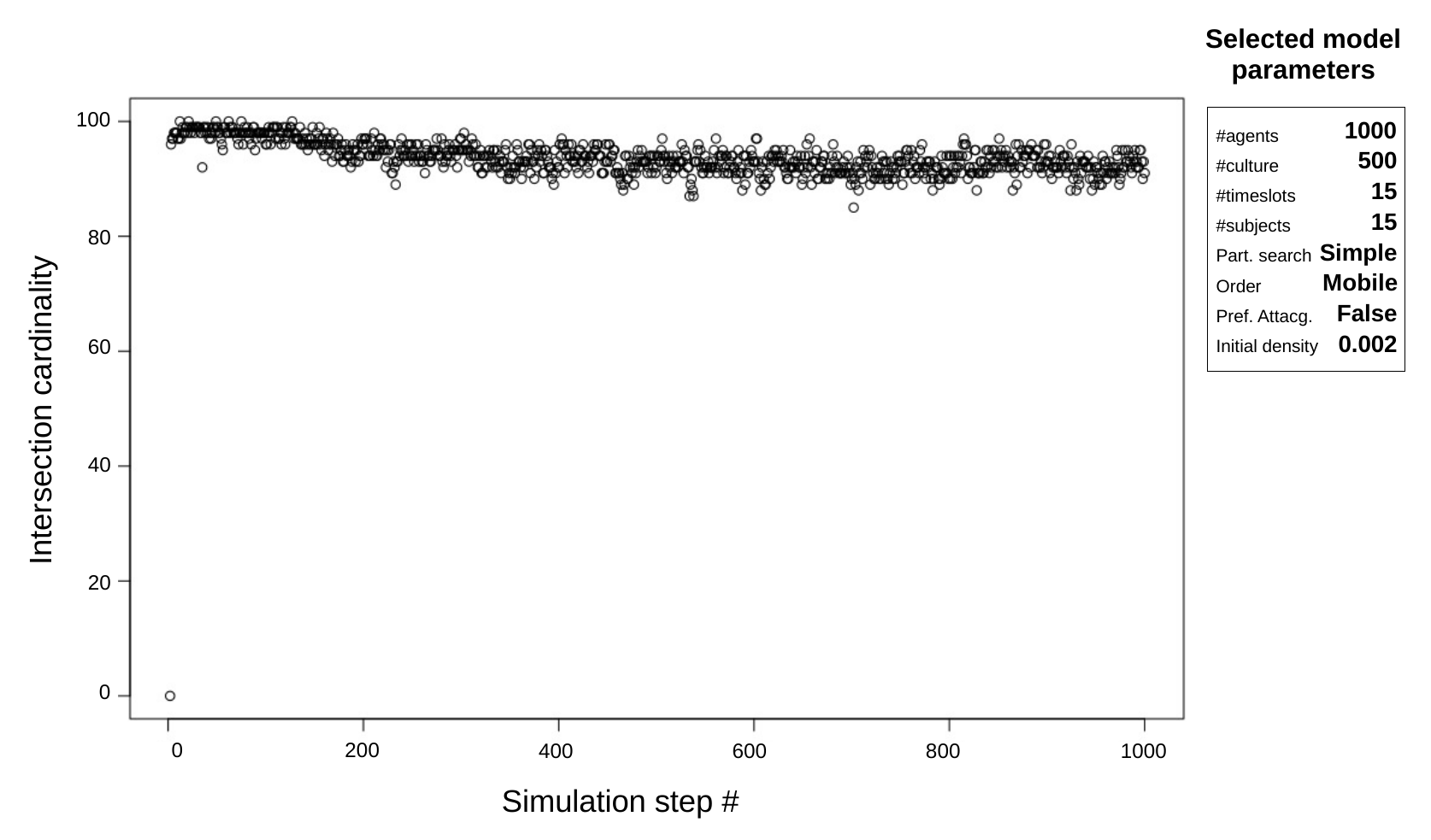}
  \end{center}
  
  \caption{\small Intersections cardinality for the entire model run for a sample of parameter sets, slight downward trend – rigidity loosening.}
  \label{fig11}
\end{figure}

\begin{figure}[!b]
  \begin{center}
    \includegraphics[width=4.3in]{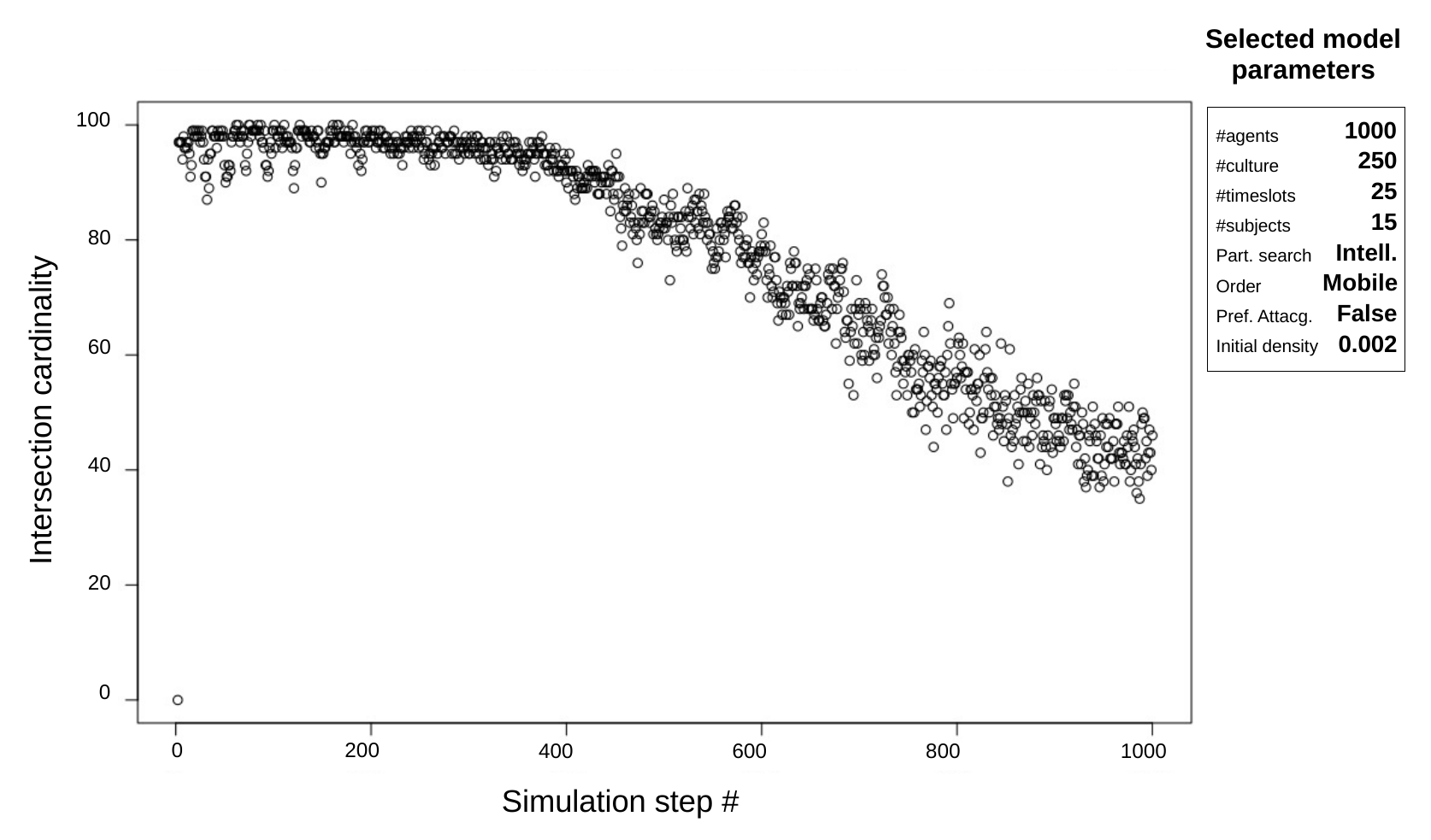}
  \end{center}
  
  \caption{\small Intersections cardinality for the entire model run for a sample of parameter sets, sudden plummeting in the last model phase on the right.}
  \label{fig12}
\end{figure}

The steady decrease in intersection values (see Fig. \ref{fig11}) is common, and we assume it is a sign of progressive loosening of the system rigidity for the Mobile execution method. However, there are also cases in which this decrease speeds up significantly, as is visible in Figure \ref{fig12}. Around step numbers 300 to 400, the intersection cardinality might plummet. This effect is tightly linked to the number of available time resources. The more weekdays available, the more cases are affected with the sharp decline in the last 600 to 700 steps of the model runs, as shown in Tables \ref{table3a} and \ref{table3b}.

\begin{table}[!ht]
\caption{
{\bf Summary of the intersection cardinality vs. step number correlation for different parameter sets. Percentage of runs for given set of parameters for which correlation is within one of three ranges.}}
\begin{tabular}{cc|c|c|c|}
\cline{3-5}
 & & \multicolumn{3}{|c|}{\bf from 500th to 1000th step}\\ \cline{1-5}
\multicolumn{2}{ |c| }{ranges of correlation value} & $\langle-1;-0.1)$ &	$\langle-0.1;0.1\rangle$ &	$(0.1;1\rangle$\\ \hline \hline
\multicolumn{1}{ |c| }{\multirow{2}{*}{\textbf{Culture size}}}	& 250	& 52.78\% &	41.67\% & 5.56\% \\ \cline{2-5}
\multicolumn{1}{ |c| }{} & 500 & 52.78\% &	41.67\% &	5.56\%\\ \hline \hline
\multicolumn{1}{ |c| }{\textbf{Number of}} & 5 &	20.83\% &	75.00\% &	4.17\% \\ \cline{2-5}
\multicolumn{1}{ |c| }{\textbf{weekdays}} & 15 &	54.17\% &	41.67\% &	4.17\% \\ \cline{2-5}
\multicolumn{1}{ |c| }{\textbf{(timeslots)}} & 25 &	83.33\% &	8.33\% &	8.33\%\\ \hline \hline
\multicolumn{1}{ |c| }{\textbf{New links}} & 	Preferential &	50.00\% &	44.44\% &	5.56\%\\ \cline{2-5}
\multicolumn{1}{ |c| }{\textbf{algorithm}} & 	Uniform &	55.56\% &	38.89\% &	5.56\% \\ \hline \hline
\multicolumn{1}{ |c| }{\multirow{2}{*}{\textbf{Arrangements}}} & Intelligent	& 52.78\% &	41.67\% &	5.56\% \\ \cline{2-5}
\multicolumn{1}{ |c| }{} & Simple & 52.78\% &	41.67\% &	5.56\% \\ \hline \hline
\multicolumn{1}{ |c| }{\multirow{3}{*}{\textbf{Relations update}}} & 1850 &	62.50\% &	33.33\% &	4.17\% \\ \cline{2-5}
\multicolumn{1}{ |c| }{} & 5018 &	33.33\% &	58.33\% &	8.33\% \\ \cline{2-5}
\multicolumn{1}{ |c| }{} & 5050	& 62.50\% &	33.33\% &	4.17\% \\ \hline \cline{3-5}
& & \cellcolor{lightgray}\textbf{52.78}\% & \cellcolor{lightgray}\textbf{41.67\%} &	\cellcolor{lightgray}\textbf{5.56}\%\\ \cline{3-5}
\end{tabular}
\label{table3a}
\end{table}

\begin{table}[!ht]
\caption{
{\bf Summary of the intersection cardinality vs. step number correlation for different parameter sets. Percentage of runs for given set of parameters for which correlation is within one of three ranges.}}
\begin{tabular}{cc|c|c|c|}
\cline{3-5}
 & & \multicolumn{3}{|c|}{\bf from 500th to 1000th step}\\ \cline{1-5}
\multicolumn{2}{ |c| }{ranges of correlation value} & $\langle-1;-0.1)$ &	$\langle-0.1;0.1\rangle$ &	$(0.1;1\rangle$\\ \hline \hline
\multicolumn{1}{ |c| }{\multirow{2}{*}{\textbf{Culture size}}}	& 250	& 77.78\% &	16.67\% & 5.56\% \\ \cline{2-5}
\multicolumn{1}{ |c| }{} & 500 & 75.00\% &	11.11\% &	13.89\%\\ \hline \hline
\multicolumn{1}{ |c| }{\textbf{Number of}} & 5 &	66.67\% &	20.83\% &	12.50\% \\ \cline{2-5}
\multicolumn{1}{ |c| }{\textbf{weekdays}} & 15 &	87.50\% &	4.17\% &	8.33\% \\ \cline{2-5}
\multicolumn{1}{ |c| }{\textbf{(timeslots)}} & 25 &	75.00\% &	16.67\% &	8.33\%\\ \hline \hline
\multicolumn{1}{ |c| }{\textbf{New links}} & 	Preferential &	72.22\% &	16.67\% &	11.11\%\\ \cline{2-5}
\multicolumn{1}{ |c| }{\textbf{algorithm}} & 	Uniform &	80.56\% &	11.11\% &	8.33\% \\ \hline \hline
\multicolumn{1}{ |c| }{\multirow{2}{*}{\textbf{Arrangements}}} & Intelligent	& 75.00\% &	11.11\% &	13.89\% \\ \cline{2-5}
\multicolumn{1}{ |c| }{} & Simple & 77.78\% &	16.67\% &	5.56\% \\ \hline \hline
\multicolumn{1}{ |c| }{\multirow{3}{*}{\textbf{Relations update}}} & 1850 &	79.17\% &	12.50\% &	8.33\% \\ \cline{2-5}
\multicolumn{1}{ |c| }{} & 5018 &	62.50\% &	16.67\% &	20.83\% \\ \cline{2-5}
\multicolumn{1}{ |c| }{} & 5050	& 87.50\% &	12.50\% &	0.00\% \\ \hline \cline{3-5}
& & \cellcolor{lightgray}\textbf{76.39}\% & \cellcolor{lightgray}\textbf{13.89\%} &	\cellcolor{lightgray}\textbf{9.72}\%\\ \cline{3-5}
\end{tabular}
\label{table3b}
\end{table}

\subsection*{Preferential attachment strengthens the emerging hierarchy}

Preferential attachment, as one of the possible heuristics for creating new edges, was implemented in the model, as it can be expected to make the effect of emergent hierarchy more visible. No significant differences in willingness usage or overall efficiency were found between model runs with preferential attachment edge building vs. non-preferential building, but there was one particularly interesting relation. 

Results for preferential attachment procedure are collected in Table \ref{table4}. For all model runs with preferential attachment turned on, the average Kendall Tau distance value amounts to 0.471, while the model runs with non-preferential attachment during new edge building resulted in distance average values of 0.476. We can conclude that introduction of preferential attachment to the model results in bringing more structure to the order in which the agents are being executed. This is on the condition that we consider only the Mobile execution method, that is, execution in the order of weighted node degree. We also examined the difference in Kendall Tau distance for preferential and non-preferential new edge building for each single parameter set, together with the significance of this difference. 

It turns out that in nearly 81\% of cases, not only did preferential attachment show lower values of Kendall Tau distance, but also the difference was statistically significant. Moreover, this scheme is also linked with the number of time resources available in the model. The percentage of cases in which preferential attachment results in significantly lower distance grows from 50\% to 100\% with the number of weekdays available in the agent’s schedule. We conclude that preferential attachment as a mechanism of building new edges in the graphs not only prompts the emergence of the hierarchy but also is tightly dependent on the resources. Observed effect is statistically significant.

\begin{table}[!ht]
\caption{
{\bf Percentage of runs for which Kendall-Tau distance is lower when preferential attachment algorithm for creating new edges is used.}}
\begin{tabular}{cc|c|}
\cline{3-3}
 & & \% of runs\\ \hline
\multicolumn{1}{ |c| }{\multirow{2}{*}{Culture size}}	& 250	& 83.33\% \\ \cline{2-3}
\multicolumn{1}{ |c| }{} & 500 & 77.78\% \\ \hline \hline
\multicolumn{1}{ |c| }{Number of} & 5 & 50.00\%	\\ \cline{2-3}
\multicolumn{1}{ |c| }{weekdays} & 15 &	91.67\% \\ \cline{2-3}
\multicolumn{1}{ |c| }{(timeslots)} & 25 &	100.00\% \\ \hline \hline
\multicolumn{1}{ |c| }{\multirow{2}{*}{Arrangements}} & Intelligent	& 83.33\%\\ \cline{2-3}
\multicolumn{1}{ |c| }{} & Simple & 77.78\% \\ \hline \hline
\multicolumn{1}{ |c| }{\multirow{3}{*}{Relations update}} & 1850	& 91.67\%\\ \cline{2-3}
\multicolumn{1}{ |c| }{} & 5018 & 66.67\% \\ \cline{2-3}
\multicolumn{1}{ |c| }{} & 5050 & 83.3\% \\ \hline \cline{3-3}
& & \cellcolor{lightgray}\textbf{80.56\%}\\ \cline{3-3}
\end{tabular}
\label{table4}
\end{table}

\section*{Discussion}
Our metaphor for an egalitarian system, with equal access to time resources, showed the best long-term performance. For a longer simulation run, the superiority of the egalitarian agent execution order over hierarchical order in terms of global effectiveness (average Willingness Usage values) is more visible, reaching 100\% of cases when considering the last simulation step (1,000th). Moreover the performance of the egalitarian system is inversely proportional to the number of available resources, reaching the best efficiency when the time is scarce. The obtained results strongly support the first hypothesis (H1) for a broad range of different scenarios and parameters.

Since, in the long run, an egalitarian system performs better anyway, we can only expect that the intelligent method of arranging meetings mentioned in the second hypothesis (H2) will have the best performance for the initial phase of simulation. Indeed, intelligent methods, i.e., mimicking the use of modern technology to increase the probability of finding a valuable partner and utilizing more of the agent’s preferences, show greater values of average willingness usage, but this effect is temporary and occurs in the initial phase of the simulation. This effect is the strongest observed for societies stratified by weighted node degree (i.e. \textit{mobile method}). On the other hand, for the hierarchical execution order, intelligent time management methods seem to contribute most to an improvement of global effectiveness. 

To verify the third hypothesis (H3), we have used the implemented Mobile execution order, which was referred to as a stratified system with added mobility for agents to move on the hierarchy ladder. We attempted to determine whether, despite some degree of the flexibility, the hierarchy can arise and sustain. The results show that the hierarchy can indeed be emergent but only in very specific condition and in the short term. The Mobile model running with resources set to high availability yields signs of a crystalizing hierarchy in the short term. Until approximately half of a single simulation run, the order in which the agents are executed stabilizes. This tendency is strong when the resources are in plentiful supply and the preferential attachment mode is enabled. However, in the long term this effect does not persist, and the hierarchy disappears, as the agents are using the enabled mobility in the model. Therefore, the third hypothesis was also confirmed. 


All three hypotheses raised in the introduction have been confirmed. Sensitivity analysis shows that the obtained results are quite stable for a broad range of parameter combinations. Social stratification, as an emergent property of a model, can itself be a broad topic for research and an innovative means of comprehending the existence of inequality in real systems. Moreover, the developed model provides an opportunity to study methods for overcoming social stratification, such as asynchronous methods of communication (e.g., publication of agents’ calendars and current interests).

The results presented in this paper do not provide an ultimate answer to the question of whether a particular level of stratification is optimal for society in general. However, obtained results shed light on the reason for the decreased effectiveness and long-term stability of hierarchies in the knowledge society. This reason is that hierarchies decrease the global effectiveness of time management, measured by the degree to which agents manage to exchange ideas and discuss with each other. Our model, based on a relatively simple process of making appointments to discuss various topics by agents embedded in an evolving social network, is quite suitable to advanced knowledge societies such as the Silicon Valley. The model is also able to shed insights on the emergent properties of an advanced knowledge society. While there may be many ways to improve the model and generalize or validate our findings, we consider them interesting areas for further study, enabled by the contributions of this article to the research community.

\bibliographystyle{alpha}
\bibliography{sample}

\newpage
\section*{Appendix I}

\begin{figure}[!b]
  \begin{center}
    \includegraphics[width=2.8in]{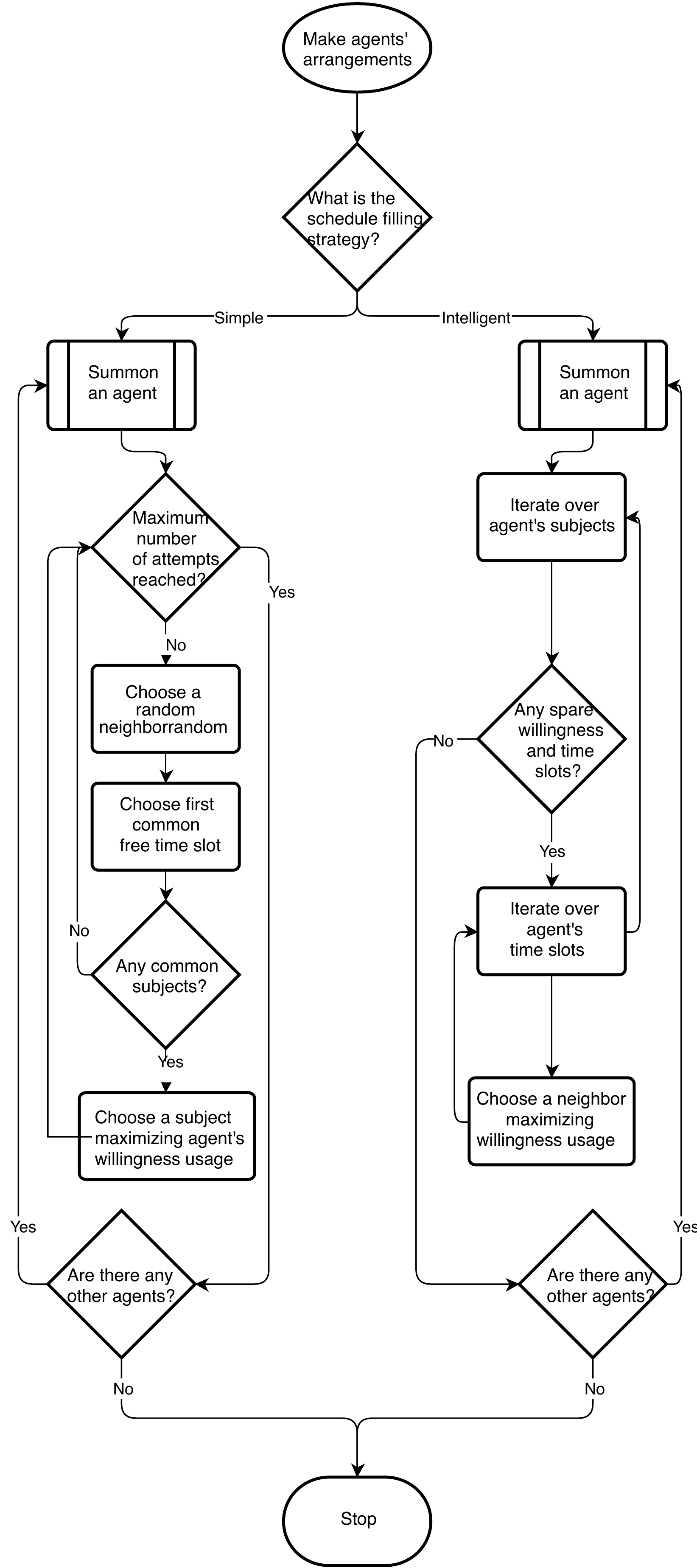}
  \end{center}
  
  \caption{\small Simple and Intelligent schedule-filling procedure flowcharts.}
  \label{appendix:fig1}
\end{figure}

\end{document}